\documentclass[journal,two side,web]{ieeecolor}
\usepackage{generic}
\newif\ifdraft
\draftfalse

\usepackage{cite}
\usepackage{amsmath,amssymb,amsfonts}
\usepackage{algorithmic}
\usepackage{graphicx}
\usepackage{algorithm,algorithmic}
\usepackage{hyperref}
\usepackage{amsmath}
\usepackage{lipsum}
\usepackage{multirow} 
\hypersetup{hidelinks=true}
\usepackage{textcomp}
\def\BibTeX{{\rm B\kern-.05em{\sc i\kern-.025em b}\kern-.08em
    T\kern-.1667em\lower.7ex\hbox{E}\kern-.125emX}}
\newcommand{\heading}[1]{\noindent\textbf{#1}}

\usepackage[normalem]{ulem} 
\newcommand{\stkout}[1]{\ifmmode\text{\sout{\ensuremath{#1}}}\else\sout{#1}\fi}

\ifdraft

\newcommand{\deleted}[1]{\textcolor{red}{\stkout{#1}}}

\newcommand{\deletedfloat}[1]{}
\newcommand{\commented}[1]{\textcolor{blue}{#1}}
\else

\newcommand{\deleted}[1]{}

\newcommand{\deletedfloat}[1]{}
\newcommand{\commented}[1]{}
\fi

\begin{document}
\title{Benchmarking machine learning for bowel sound pattern classification -- from tabular features to pretrained models}

\author{Zahra Mansour, Verena Uslar, Dirk Weyhe, Danilo Hollosi and Nils Strodthoff.
\thanks{Zahra Mansour, Verena Uslar, Dirk Weyhe, and Nils Strodthoff are with the University of Oldenburg, 26129 Oldenburg, Germany (e-mail: firstname.lastname@uol.de). Zahra Mansour is co-affiliated with Fraunhofer IDMT, an institute part of HSA. Danilo Hollosi is with Fraunhofer IDMT, institute part HSA (e-mail: danilo.hollosi@idmt.fraunhofer.de) Corresponding author: Nils Strodthoff}}

\maketitle

\begin{abstract}
The development of electronic stethoscopes and wearable recording sensors opened the door to the automated analysis of bowel sound (BS) signals. This enables a data-driven analysis of bowel sound patterns, their interrelations, and their correlation to different pathologies. This work leverages a BS dataset collected from 16 healthy subjects that was annotated according to four established BS patterns.
This dataset is used to evaluate the performance of machine learning models to detect and/or classify BS patterns. The selection of considered models covers models using tabular features, convolutional neural networks based on spectrograms and models pre-trained on large audio datasets. The results highlight the clear superiority of pre-trained models, particularly in detecting classes with few samples, achieving an AUC of 0.89 in distinguishing BS from non-BS using a HuBERT model and an AUC of 0.89 in differentiating bowel sound patterns using a Wav2Vec 2.0 model.
These results pave the way for an improved understanding of bowel sounds in general and future machine-learning-driven diagnostic applications for gastrointestinal examinations.
\end{abstract}


\section{Introduction}
\label{sec:introduction}
\IEEEPARstart{B}{owel} sound auscultation represents an essential non-invasive physical examination technique. Although it is part of the abdominal examination routine worldwide, its medical value remains limited and subjective \cite{felder2014}. This shortcoming mainly concerns the unspecified nature of the bowel sound (BS) signal, its irregular occurrence with comparably long typical time frames, and its dependency on the diet. These aspects make it challenging to identify and characterize the typical waveform or pattern of the BS signal\cite{nowak2021}.

\heading{History of BS analysis} Analysing the BS was first introduced in the early twentieth century by Cannon\cite{cannon1905}, who described two types of BS: rhythmic sounds related to the movement of the intestine and continuous random sounds found at different locations within the bowel\cite{cannon1905}. 
Since human auscultation depends on the expertise of the clinician, it cannot reliably distinguish between normal and pathological BS \cite{felder2014}, therefore following studies tended to automate the analysis of the BS by extracting time domain-related features \cite{dalle1975}, such as amplitude, duration, signal to signal interval and total number of sounds per minute \cite{dalle1975,bray1997,ranta2004,ranta2010}, aiming to discriminate how those features might vary in the case of some diseases such as irritable bowel syndrome (IBS) \cite{craine1999,craine2001} \cite{craine2002,hadjileontiadis1999}, obstruction \cite{yoshino1990,ching2012} and post-operative illus \cite{kaneshiro2016,spiegel2014}.

\heading{Automatic BS analysis} The scope of BS analysis ranges from simply enhancing the quality of the recorded BS\cite{hadjileontiadis1999}, to detect BS events in the abdominal recording\cite{Baronetto2023}, and in some cases to identify different kinds of BS patterns\cite{dimoulas2007}.

\heading{Research gap} Despite the progress in BS signal detection and pattern recognition, there remains a significant research gap. While deep learning (DL) and transfer learning models have revolutionized audio classification tasks, particularly for heart and lung sounds, their application to BS analysis remains limited. Only a few recent studies have employed DL models to detect BS signals \cite{Kutsumi2023, Zhao2020, Zhao2022}. However, no study has yet explored the use of DL or transfer learning for classifying BS patterns. Addressing this gap could unlock new insights into the relationship between BS patterns and gastrointestinal function or pathology.

\begin{figure*}[ht]
\centering
\includegraphics[width=.8\textwidth]{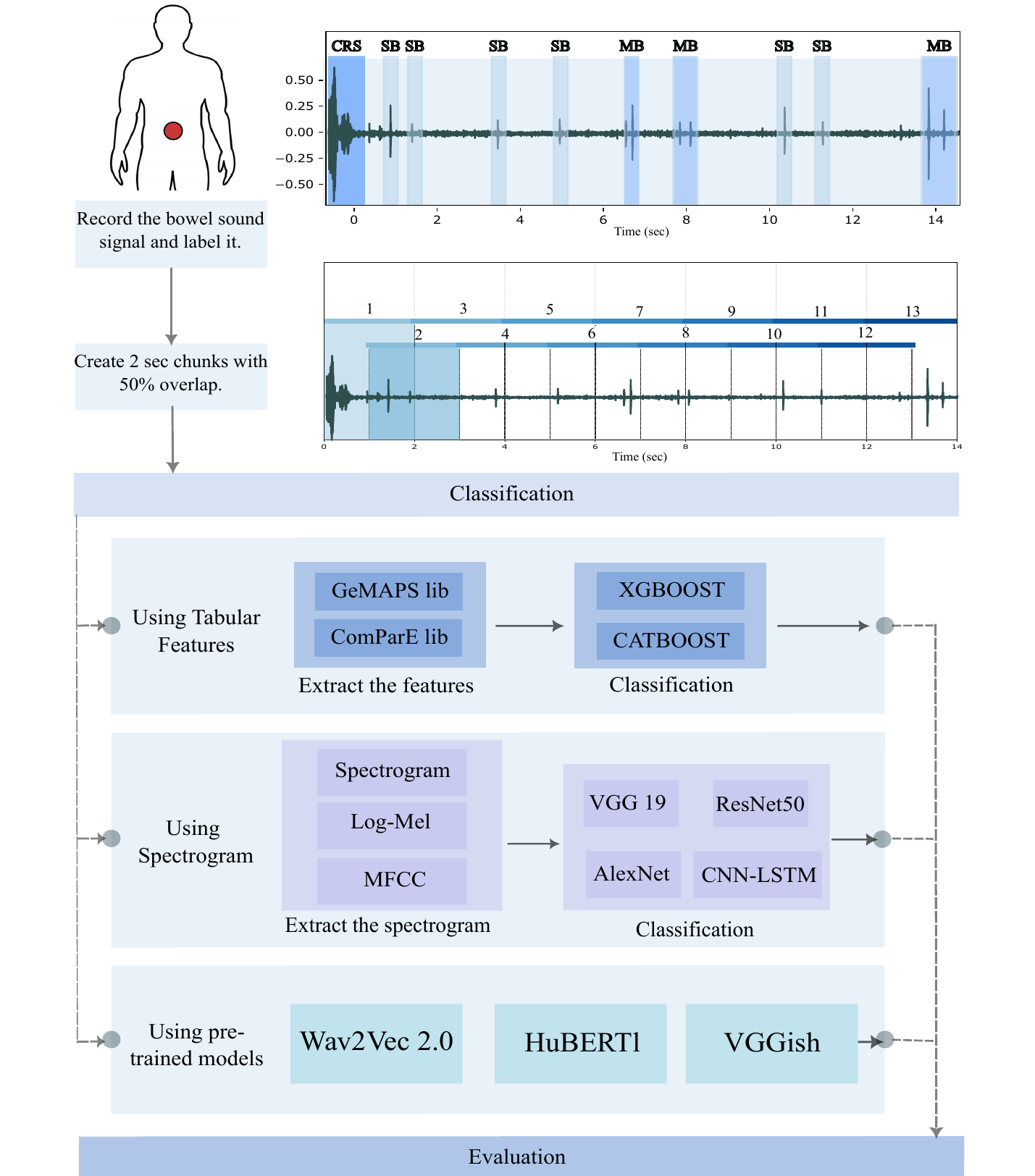}
\caption{Schematic diagram summarizing the course of the study. It starts by recording the BS by placing the sensor on the subject abdomen, proceeds over labeling the signal into non-BS and  BS patterns (SB, MB, CRS, HS), and segmenting the signal into 2 seconds overlapped windows, to use it later on the classification with 3 different methods (using tabular features, using spectrogram and using pre-trained models).}
\label{fig1}
\end{figure*}

\heading{Summary of this study} In this study we compare the performance between tabular features extraction models, deep learning models (VGG 19, ResNet 50, and AlexNet) with different spectrograms as input (Standard, Log-Mel, and MFCC), and transfer learning based models (Wav2Vec 2.0, HuBERT and VGGish). In two different tasks, distinguish between non-BS and BS signals, and classify between non-BS signals and the BS patterns(SB, MB, CRS, and HS) The methodological framework underlying this study is summarized in Figure~\ref{fig1}. 

\begin{figure*}[!t]
\centerline{\includegraphics[width=\textwidth]{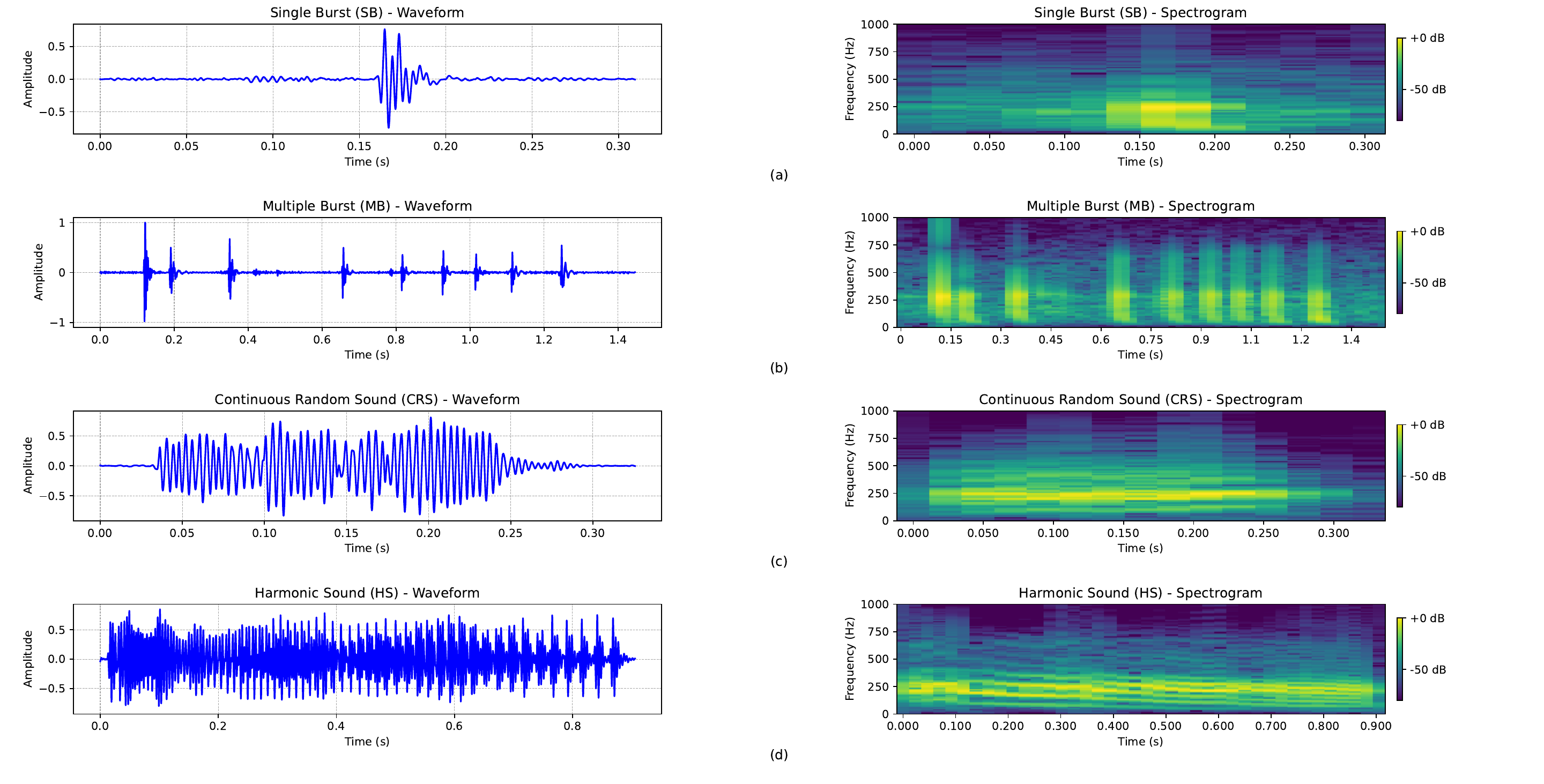}}
\caption{Bowel sound patterns examples extracted from the dataset used in this study, the left column represents the signal on the time domain, and the right column describes the signal in the frequency domain. Starting from the top, (a) Single Burst (SB), (b) Multiple Burst, (c) Continuous Random sound (CRS), and (d) Harmonic Sound (HS).}
\label{fig2}
\end{figure*}

\section{Materials and Methods}
\subsection{Background}
\heading{Physiological origin} It has been established that two sources mainly produce BS: the contractions and relaxations of the abdominal wall smooth muscle which leads to pushing the intestines contents through the gastrointestinal (GI) tract, and the interaction of the gas and the partially digested food through the intestines \cite{allwood2019}. This explanation has been ensured by electromyography and anatomy research, which showed that the intestine contraction activity appears as a burst or explosive peak (very short contraction) or as clustered contractions with regular release and slope duration \cite{nowak2021}. 
In light of this explanation, bowel sounds' most observed patterns can be divided into two main groups: quick clicks that could appear as a single or sequence of bursts and longer contractions with rubbing or piping noises.

\heading{Investigated BS patterns} Therefore, the bowel sound patterns investigated in this study combine the types described within the abdominal auscultation medical terms in addition to the bowel sound patterns that have been mentioned in the previous studies \cite{du2018} \cite{dimoulas2006} \cite{wang2022}. The bowel sound patterns in the time and frequency domain are presented in Figure~\ref{fig2} and can be categorized into four groups: 

\begin{itemize}
\item \heading{Single Burst (SB)/Solitary Clicks:} SB is the most frequent bowel sound type, it has a damped sinusoidal-like nature and is shown as pulses with significantly reduced energy. SB is likely caused by small GI contractions or splashes of the digestive content. It is mostly characterized by very short duration ranges between 10–30 ms, with a frequency peak around 400 Hz Figure~\ref{fig2}a.
\item \heading{Multiple Burst (MB)/Repeated Clicks:}  MB is represented by multiple SBs with a short inconsistent silent gap between adjacent components. It might be produced by mixing the food and gas inside the intestine. Each component in the MB spectrogram looks quite similar with slight differences in bandwidth and amplitude. MB is longer than SB with duration ranges from 40 to 1500 ms Figure~\ref{fig2}b. 
\item \heading{Continuous Random Sound (CRS)/Crepitating Sweeps:} CRS contains a main clustered contraction, sounding like crepitating or rumbling, which is probably caused by pushing the fluid and gas through the intestine. CRS has a comparably longer duration from 200 ms to 4000 ms without any noticeable silent gaps, and a higher frequency range 500-1700 Hz, see Figure~\ref{fig2}c. 
\item \heading{Harmonic Sound (HS)/ Whistling Sweeps:} This is the least frequent bowel sound pattern, composed of three to four clear frequency components that appear in the spectrogram which cause piping notes with a whistling-like sound. The formation of intestinal stenosis, or gastric activity produces it. HS duration ranges from 50 ms to 1500 ms, while HS spectrogram contains peak frequencies as multiples of the fundamental frequency, which is usually relatively low, around 200 Hz, see Figure~\ref{fig2}.d.
\end{itemize}

\subsection{Related work}
\heading{Signal processing} The majority of studies investigating bowel sounds primarily focus on extracting and enhancing the BS signal using advanced signal processing techniques. Examples include the Wavelet Transform-Based Stationary-Non-Stationary Filter \cite{hadjileontiadis1999,liatsos2003}, Short Time Fourier Transformation \cite{ficek2021}, and Fractal Dimension \cite{dimoulas2007}. Similarly, machine learning techniques, such as the use of jitter and shimmer parameters \cite{kim2011}, neural networks \cite{yin2015,kim22011}, and support vector machines \cite{yin2018}, have been employed for BS signal processing and analysis.

\heading{BS identification and classification}While these methods provide robust tools for signal detection and enhancement, only a limited number of studies have focused on identifying and classifying BS patterns. The early work by Dimoulas et al \cite{dimoulas2003} introduced abdominal sound pattern analysis (ASPA), combining BS time-energy alterations with electrical or pressure signals. This study marked the first step toward understanding BS patterns in the context of gastrointestinal function.

Building on these foundations, further research aimed to autonomously classify BS patterns. For instance, one study employed a wavelet neural network to differentiate between two BS patterns and three types of interfering noises \cite{dimoulas2007}. Another significant study analyzed two hours of BS recordings from ten participants, identifying five distinct BS types based on waveform and spectrogram characteristics. This study also investigated inter-subject variations in the duration and proportion of these patterns \cite{du2018}. Similarly, research involving 1,140 BS recordings from 15 volunteers identified four unique BS patterns \cite{Zhang2024}.

More recently, advancements in deep learning have facilitated new approaches to BS analysis. A convolutional neural network (CNN)-based detector was developed to identify four types of BS and to investigate the acoustic effects of food consumption. This study proposed that total BS duration increases post-consumption, highlighting the relevance of studying BS patterns in relation to the digestion cycle \cite{wang2022}.
Pre-trained models have been introduced for the first time in BS identification by \cite{Baronetto2023}, where different pre-trained DNNs have been developed for BS event spotting. To the best of our knowledge, no prior work has examined transfer learning strategies for BS pattern classification.

\subsection{Dataset}
\heading{Dataset collection} The dataset for this study was acquired by recording bowel sounds from 16 healthy subjects (8 male, 8 female) with an average age of 28.6 $\pm$ 5.5 years at PIUS Hospital in Oldenburg, Germany. The recordings were made at a sampling rate of 44.5 kHz, resulting in a total of 270 minutes of acoustic data. The recordings were captured using 3M Littmann Digital CORE, Thinklabs One Digital Stethoscope, and SonicGuard sensor \cite{Mansour2024}. The three mentioned sensors were positioned on the umbilical region at the abdominal wall of the subject, using a standard stethoscope placement protocol to ensure high-quality signal acquisition. 

\begin{figure}[!t]
\centerline{\includegraphics[width=\columnwidth]{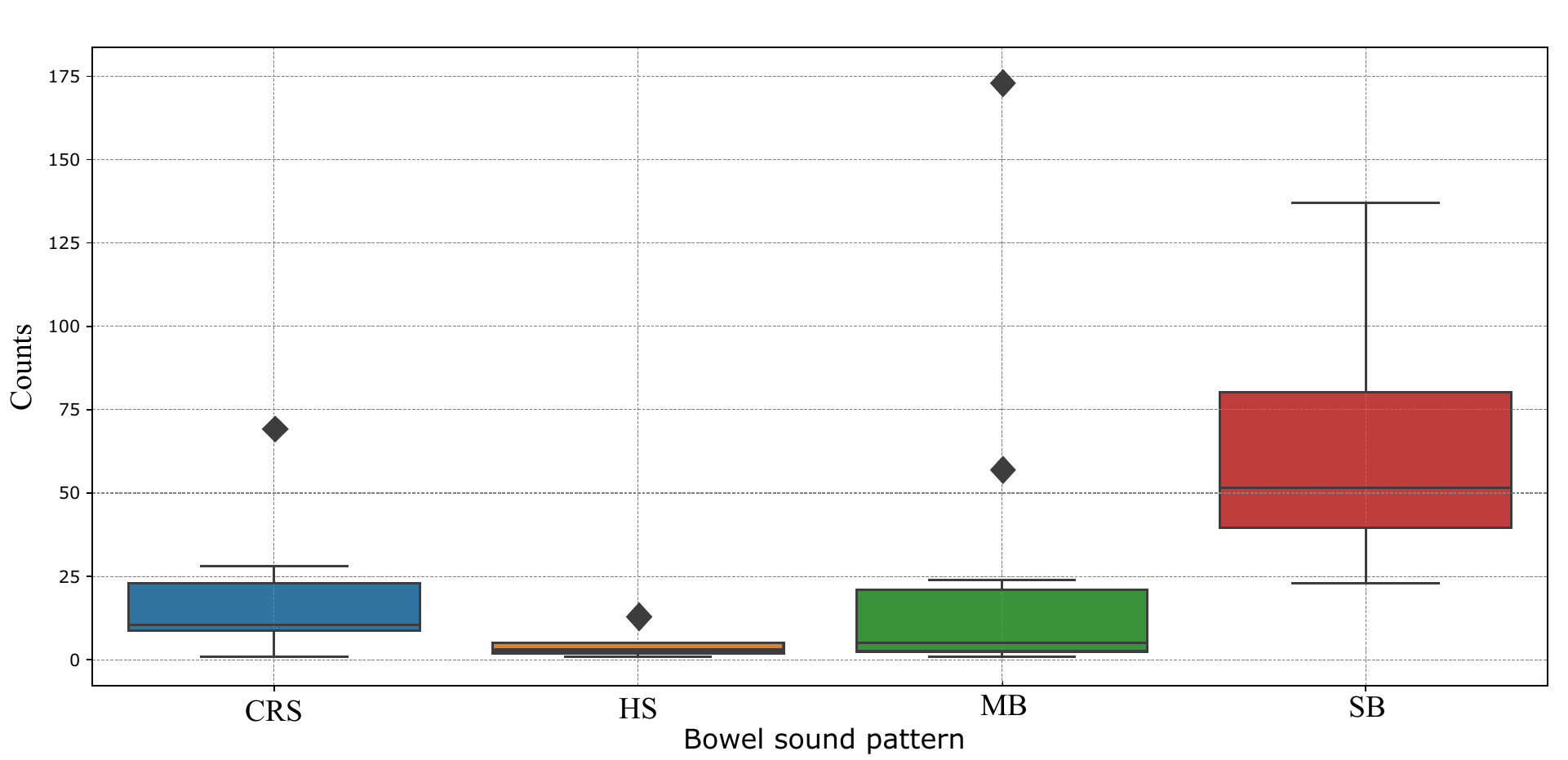}}
\caption{Distribution of bowel sound pattern(SB, MB, CRS, HS) counts by subjects, the box represents the interquartile range (IQR), with the horizontal line inside the box indicating the median. Whiskers extend to 1.5 × IQR, and points outside the whiskers represent outliers. The SB group shows the highest median and variability, while the HS group has the smallest counts and minimal variability .}
\label{fig3}
\end{figure}

\heading{Annotation} The recorded data were manually labeled into non-BS signals and BS signals. The BS signals were further subdivided into bowel sound patterns: SB, MB, CRS, and HS. The distribution of the classes in the dataset showed that SB is the most frequent BS class, representing 58.53\% of the BS signals, followed by MB (21.78\%) and CRS (17.24\%), while the least common class is HS, which accounted for only 2.45\% of the events. This distribution is consistent on a subject level, as represented in Figure~\ref{fig3}, where SB patterns are present in all subjects' signals. Similarly, MB and CRS are observed, though to a lesser extent, while HS appears in only 56\% of the subjects.

\begin{figure*}[]
\centerline{\includegraphics[width=\textwidth]{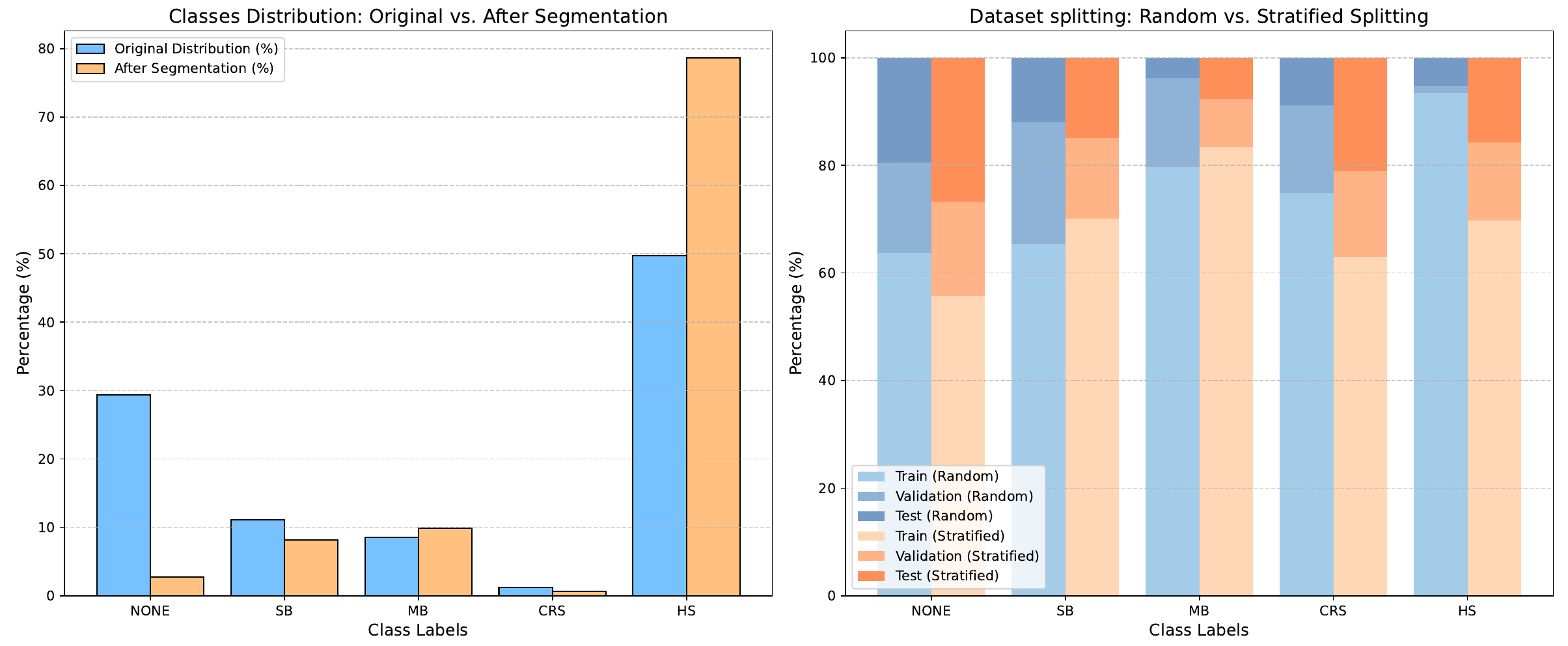}}
\caption{The first column shows the distribution of the classes (non-BS, SB, MB, HS) within the dataset before and after the segmentation using 2 2-second, overlapping window. The second column shows the distribution of the 5 classes within the train, validation, and test sets by using random and stratified split. The distribution of the classes is closer to the required ratio (70\%, 15\%, 15\%) using stratified splitting.}
\label{fig4}
\end{figure*}

\heading{Dataset preparation}
After collecting the BS and annotating it according to the BS patterns, each wave signal was divided into smaller chunks by applying a sliding window with a 2-second duration and a stride of 1s, i.e., with 50\% overlap. The BS pattern of each segment was annotated as the dominant pattern within the segment. The distribution of the classes within the dataset before and after the segmentation is represented in the first column in Fig.~\ref{fig4}. The amount of SB samples intensely dropped because of the segmentation while the non-BS class samples increased, this could be attributed to the short duration of SB, which is smaller than the segmentation window (2 sec), making it less likely for SB to be the dominant event. 

\heading{Dataset splits} 
Before building the model, the dataset is split into the train, validation, and test set (70\%, 15\%, 15\%) respectively. Because of the unequal distribution of classes within the dataset, performing random splitting produces non-homogeneous class distributions across the three sets.
Thus, stratified splitting is applied to produce samples where the proportion of the groups is maintained. To this end, we use the stratified splitting procedure from \cite{wagner2020ptb, Sechidis2011}, which supports multi-label stratification while respecting patient assignments. The result of the stratification process is visualized in Figure~\ref{fig4}.

\subsection{Considered classification models}
The process undertaken in this study is visually summarized in Figure~\ref{fig1}. We roughly distinguish three approaches: Tree-based classifiers applied to tabular features, convolutional neural networks applied to spectrograms, and finetuned pre-trained models operating on spectrograms or raw waveform input.

\subsubsection{Classification using tabular features}

\heading{Feature sets} Two sets of features were used from the openSMILE (open-source Speech and Music interpretation by Large-space Extraction) toolkit. The first one is Geneva Minimalistic Acoustic Parameter Set (GeMAPS a.v01)) for Voice Research and
Affective Computing, is a set of 62 acoustic features developed to extract emotional states from speech. It contains frequency, energy, position, and temporal characteristics \cite{Eyben2010}. The second set is ComParE 2016 (Computational Paralinguistics Challenge) which is tolerated for the Computational Paralinguistics Challenge, it covers features related to phonation, articulation, and spectral shape, contains 6300 acoustic features such as Loudness, MFCCs(Mel frequency cepstral coefficients), temporal and spectral features \cite{Weninger2013}.

\heading{Gradient-boosting decision trees} Each set of features then is used separately as an input with two gradient boosting frameworks, XGBoost (Extreme Gradient Boosting) \cite{Chen2016} and  CatBoost(Categorical Boosting) \cite{Prokhorenkova2018}. XGBoost and CatBoost are used with the following hyperparameters: max depth = 7, learning rate = 0.001, and number of iterations = 50.

\subsubsection{Classification using spectrogram}
This approach relies on converting 1D raw audio signals into 2D visual data represented by the spectrogram, allowing the use of deep learning architectures from computer vision for audio classification. Three types of spectrograms are tested:
\heading{Standard spectrogram} The standard spectrogram gives a visual representation of how the frequency components of the signal evolve over time, and it is obtained by calculating the squared magnitude of the Short Time Fourier Transform (STFT) of the wave segment. The STFT is given by \cite{Rizal2022}
\begin{equation}
X(t, f) = \int_{-\infty}^{\infty} x(\tau) w(\tau - t) e^{-j2\pi f \tau} \, d\tau\,,
\end{equation}
where $w$ is hamming function applied to the $x(\tau)$ signal to divide it into smaller portions in the time domain. 
The standard spectrogram is then computed as the power at each time-frequency point, i.e., as $S(t, f) = |X(t, f)|^2$.

\heading{Log Mel-Spectrogram} The Log Mel-Spectrogram reflects how energy is distributed over frequency and time, by mimicking human auditory perception by applying a log transformation that compresses the dynamic range of the spectrogram, and highlight the lower amplitude features which makes the data more balanced for downstream tasks. Log Mel-Spectrogram is calculated by passing The power spectrogram \( |X(f, t)|^2 \)  through a series of Mel filter banks to convert the frequency scale according to the following equation \cite{Zheng2019}:
\begin{equation}
M(f, t) = \sum_{k=1}^{K} |X(k, t)|^2 \cdot H_k(f)\,, 
\end{equation}
where \( H_k(f) \) represents the triangular filters that map linear frequencies to the Mel scale.
then the log of the Mel-spectrogram is computed via
\begin{equation}
\text{Log-Mel}(f, t) = \log(1 + M(f, t))
\label{Mel2}
\end{equation}

\heading{Mel-Frequency Cepstral Coefficients (MFCC)}
The last representation is the MFCC spectrogram, which is widely used in audio processing due to its ability to capture essential spectral properties of the sound and represent it in a compact form. It is calculated by applying a Mel-spectrogram (with the same STFT and filter bank configuration) as mentioned before and then applies a discrete cosine transform (DCT) to compute the first 13 cepstral coefficients, which provides a lower-dimensional representation by returning significant information while discarding less relevant details. It is defined by the following equation \cite{Abdul2022}, 
\begin{equation}
\text{MFCC}(c, t) = \sum_{m=0}^{M-1} \log(M(f, t)) \cdot \cos \left( \frac{\pi c (2m + 1)}{2M} \right)  \,,
\label{MFCC}
\end{equation}
where $M$ is the total number of Mel frequency bins.

Each one of the mentioned spectrograms then is used separately as an input to  the following four deep-learning models:

\heading{AlexNet} The AlexNet model was introduced by Krizhevsky et al in the 2012 \cite{Krizhevsky2012}, and demonstrated the potential of deep learning in computer vision. It represents a convolutional neural network designed to learn hierarchical features from raw pixel data, consisting of eight layers; five convolutional layers with large filters, with a Rectified Linear Unit (ReLU) applied after each one of them to help the network learn more complex patterns, followed by three fully connected layers to make the predictions. 

\heading{VGG19} The VGG19  model was originally introduced by the Visual Geometry Group (VGG) at the University of Oxford in 2014\cite{Simonyan2015}, it consists of 16 layers: 13 convolutional layers for features extraction followed by 3 fully connected layers for classification. Each convolutional layer uses a small filter (3x3) with a stride of 1 to learn various spatial hierarchies of features, while the pooling layers use a 2x2 kernel, and a stride of 2 is used to reduce the spatial dimensions of the feature maps.

\heading{ResNet50} The Residual Networks architecture, proposed by He et al in  2015\cite{He2015},  introduced the concept of residual connections. Usually, ResNet starts with a 7x7 convolutional layer for feature extraction followed by a series of residual blocks. In this study, we use a ResNet50 model.
AlexNet, VGG19, and ResNet50 models that have been used in this study, were initialized with weights pre-trained on the ImageNet dataset.

\heading{CNN-LSTM} For the last model, two neural network architectures were combined: Convolutional Neural Networks (CNNs) for extracting the spatial features and Long Short-Term Memory (LSTM) networks for capturing the temporal dependencies in the data by using the memory cells, which allow the network to retain and update information over long periods \cite{bae2016acoustic}. In this study, the CNN-LSTM model consists of 2 CNN layers with a kernel size of 3, stride of 1, and padding of 1, in addition to Max-pooling applied after each convolution. They are followed by an LSTM which consists of 2 layers stacked together with 128 hidden features. The LSTM layer outputs a sequence of hidden states for each time step, and the final hidden state is then passed through the fully connected layer to produce the final classification output. 

All the mentioned models were trained for 20 epochs, with a 0.0001 learning rate, using Cross Entropy Loss and an Adam optimizer.

\subsubsection{Classification using transfer learning}
Transfer learning is a machine learning technique where a model is trained on a large dataset and learns a set of features, then it is fine-tuned and reused as a starting point for a model with the smaller but related dataset. This technique was reported to reduce computational cost and to lead to better performance in the related field of heart sound analysis, in particular for small sample sizes \cite{Koike2020}. In this study, three models that were trained on a large audio dataset are used as a feature extractor: 

\heading{Wav2Vec 2.0} Wav2Vec 2.0 \cite{Baevski2020} is a transformer model employed by a CNN to encode raw audio into features and it is trained using a contrastive loss function to perform self-supervised pre-training on unlabeled speech data. Wav2Vec masks portions of the input audio data during the training and learns to predict these masked segments based on the surrounding context.

\heading{HuBERT} (Hidden-Unit Bidirectional Encoder Representations from Transformers) is a transformer-based model, again operating on features extracted from the raw waveform data using a shallow CNN \cite{Hsu2021}. Similar to wav2Vec it uses a masking strategy, but HuBERT introduces Hidden Units by generating initial labels on the first training and then improving them from the model's output in the following training phases. 

Both models were only trained on excessively large speech datasets, but the effectiveness of the learned representations has been demonstrated for general audio classification tasks \cite{Hsu2021}

\heading{VGGish} VGGish was developed as an adaption of the VGG network to extract features from raw audio signal \cite{Gemmeke2017}. It has been trained in a supervised fashion on AudioSet, containing over 2 million human-labeled 10-second audio segments covering variant sound events. While the original VGG networks processed 2D image data, VGGish processes log-mel spectrograms of the audio signal as an input. VGGish consists of a series of convolutional layers, each followed by ReLU activations and max-pooling layers followed by fully connected layers to compress the extracted features into a 128-dimensional embedding vector.

\subsection{Evaluation}
We base the evaluation on the area under the receiver operating curve (AUC), as a commonly used ranking metric. AUC is calculated using True Positive Rate (TPR) and False Positive Rate (FPR), which makes it balance between Sensitivity and Specificity.
For binary classification, AUC is computed using the Receiver Operating Characteristic (ROC) curve, which plots the True Positive Rate (TPR) against the False Positive Rate (FPR),
\begin{equation}
\text{AUC} = \int_{0}^{1} \text{TPR}(\text{FPR}) \, d\text{FPR}
\end{equation}
TPR is given by:
\begin{equation}
\text{TPR} = \frac{\text{TP}}{\text{TP} + \text{FN}}
\end{equation}
FPR is given by:
\begin{equation}
\text{FPR} = \frac{\text{FP}}{\text{FP} + \text{TN}}   
\end{equation}
 
As we are dealing with multi-class classification, AUC is calculated for each class against all other classes (One-vs-Rest approach). Then the overall AUC is computed by averaging the AUC for all individual classes. Multi-classes AUC calculation is given by:
\begin{equation}
\text{AUC}_{\text{macro}} = \frac{1}{C} \sum_{i=1}^{C} \text{AUC}_i\,,
\end{equation}
where $C$ is the number of classes and $\text{AUC}_i$ refers to the AUC for class \( i \).

\section{Results}
\subsection{Classifying between non-BS and BS}
\begin{figure*}[]
\centerline{\includegraphics[width=\textwidth]{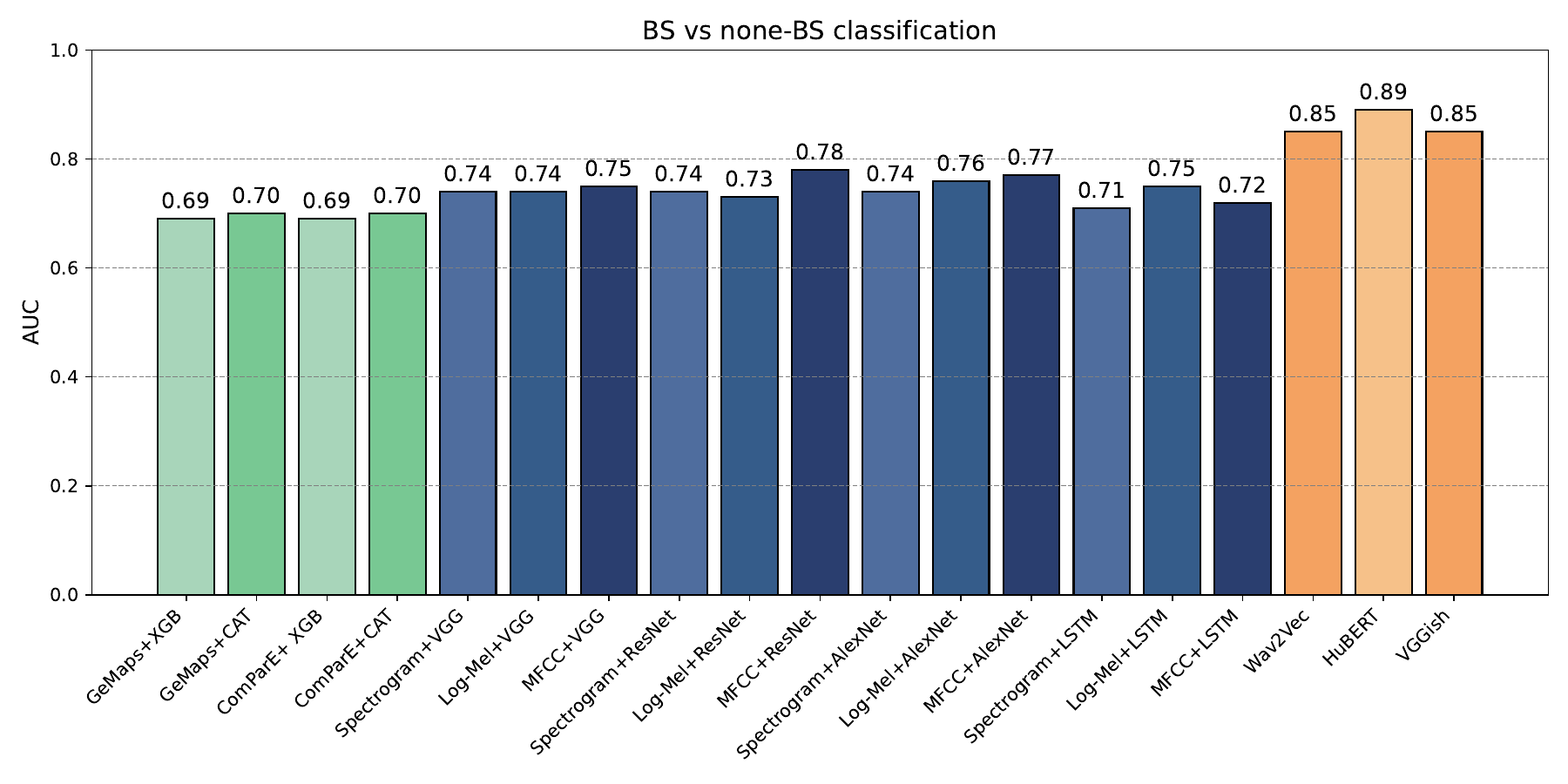}}
\caption{The AUC values of the binary classification between non-BS and BS signal are listed as follows: using tree-based models on tabular features(green; bars from 1 to 4), Spectrogram-based models (blue; bars from 5 to 16) and transfer learning based features (orange; bars from 17 to 19)}
\label{fig5}
\end{figure*}

The AUC values of the binary classification between non-BS and BS signals using the different models are listed in Figure~\ref{fig5}.

\heading{Tabular features} Using tabular input features, both feature libraries (GeMAPS and ComParE) showed similar performance, with slightly better results achieved by using CatBoost (AUC=0.7) compared to XGBoost(AUC=0.69).

\heading{Spectrograms} Turning to spectrogram-based models, all the DL models (VGG19, ResNet50, AlexNet, and CNN-LSTM) exhibit similar levels of performance with the lowest AUC obtained by using Standard spectrogram with CNN-LSTM (AUC=0.71) while the highest AUC achieved by using MFCC with ResNet50 (AUC=0.78). 
Among the spectrograms, MFCC shows better performance compared to the two other types of spectrograms in the majority of cases. 

\heading{Transfer learning} Finally concerning classification using transfer learning, all the models that have been pre-trained on a large dataset and then fine-tuned to be used for classification between BS and non-BS demonstrate the best performance among all the other models. The highest AUC is reached by using HuBERT followed by Wav2Vec2.0 and VGGish (0.89, 0.85, 0.85) respectively. This effectiveness can be attributed to the fact that the pretrained model starts with weights already optimized on a large dataset, requiring fewer adjustments during finetuning. 

\subsection{Classification between non-BS and BS patterns(SB, MB, CRS, HS)}
\begin{figure*}[]
\centerline{\includegraphics[width=\textwidth]{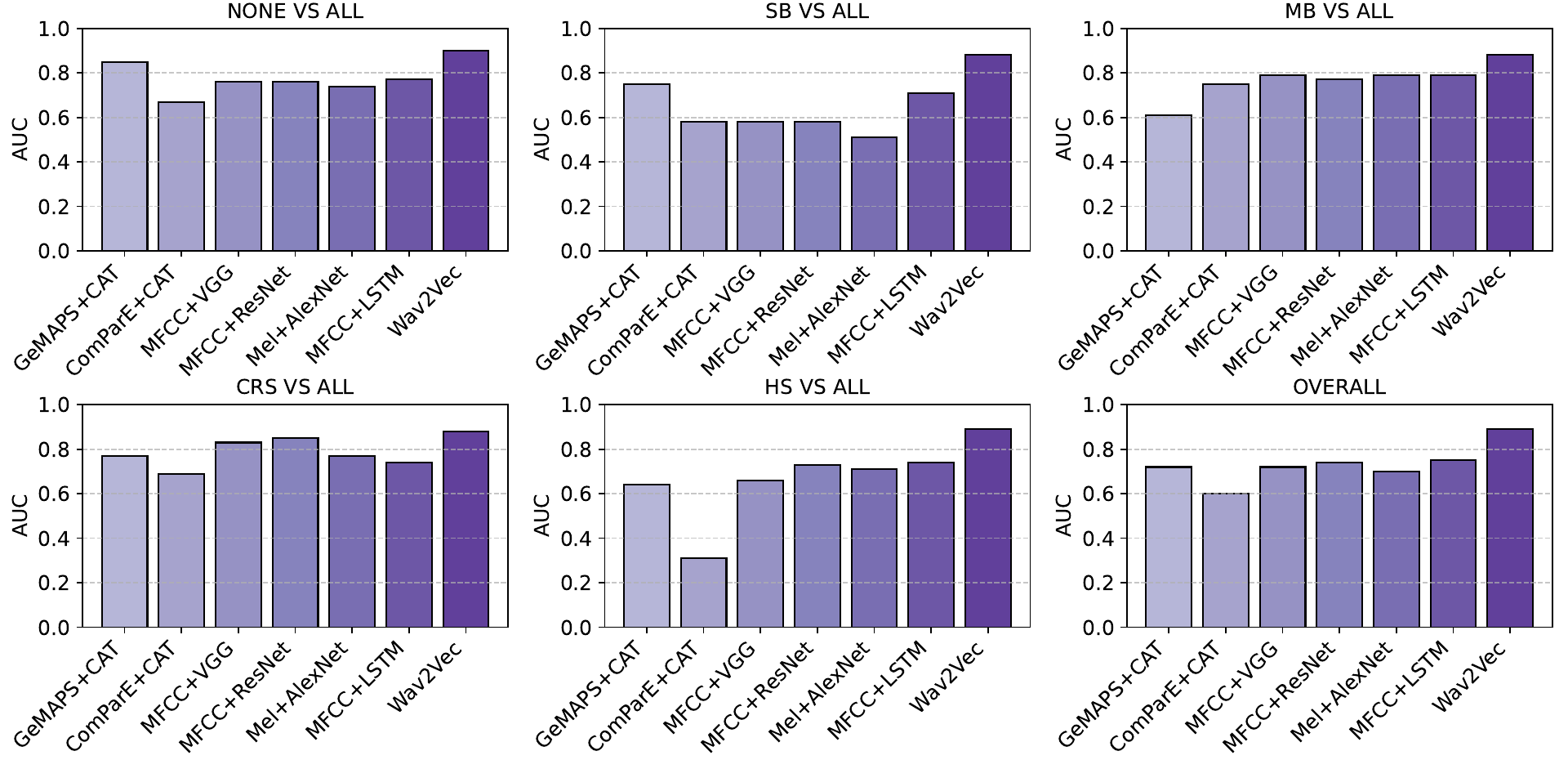}}
\caption{The binary AUC values for each class vs the rest(NONE vs all, SB vs all, MB vs all, CRS vs all, and HS vs all) and the overall AUC, achieved by the best models within each technique group.}
\label{fig6}
\end{figure*}

In this section, the results of classifying between 5 classes (non-BS, SB, MB, CRS, and HS) are presented. The AUC of each class and the overall AUC are listed in Table~\ref{table1}. A bar chart showing the AUC value for each class, along with the overall AUC of the best model from each group of classifiers is presented in Figure~\ref{fig6}.
\begin{table*}[htbp]
\caption{The binary AUC values for each class vs the rest(NONE vs all, SB vs all, MB vs all, CRS vs all, and HS vs all) and the overall AUC, using tabular features, spectrograms, and transfer learning-based models}
\label{table1}
\centering
\begin{tabular}{|c|cc|cccccc|}
\hline
\multirow{2}{*}{\textbf{}} & \multicolumn{1}{c|}{\multirow{2}{*}{\textbf{Features}}} & \multirow{2}{*}{\textbf{Model}} & \multicolumn{6}{c|}{\textbf{AUC}} \\ \cline{4-9} 
 & \multicolumn{1}{c|}{} &  & \multicolumn{1}{c|}{\textbf{None}} & \multicolumn{1}{c|}{\textbf{SB}} & \multicolumn{1}{c|}{\textbf{MB}} & \multicolumn{1}{c|}{\textbf{CRS}} & \multicolumn{1}{c|}{\textbf{HS}} & \textbf{Macro} \\ \hline
\multirow{4}{*}{\textbf{Tabular features}} & \multicolumn{1}{c|}{\textbf{GeMAPS}} & \textbf{XGB} & \multicolumn{1}{c|}{0.73} & \multicolumn{1}{c|}{0.62} & \multicolumn{1}{c|}{0.53} & \multicolumn{1}{c|}{0.83} & \multicolumn{1}{c|}{0.36} & 0.61 \\
 & \multicolumn{1}{c|}{\textbf{GeMAPS}} & \textbf{CAT} & \multicolumn{1}{c|}{0.85} & \multicolumn{1}{c|}{0.75} & \multicolumn{1}{c|}{0.61} & \multicolumn{1}{c|}{0.77} & \multicolumn{1}{c|}{0.64} & 0.72 \\ \cline{2-9} 
 & \multicolumn{1}{c|}{\textbf{ComParE}} & \textbf{XGB} & \multicolumn{1}{c|}{0.63} & \multicolumn{1}{c|}{0.45} & \multicolumn{1}{c|}{0.34} & \multicolumn{1}{c|}{0.71} & \multicolumn{1}{c|}{0.14} & 0.45 \\
 & \multicolumn{1}{c|}{\textbf{ComParE}} & \textbf{CAT} & \multicolumn{1}{c|}{0.67} & \multicolumn{1}{c|}{0.58} & \multicolumn{1}{c|}{0.75} & \multicolumn{1}{c|}{0.69} & \multicolumn{1}{c|}{0.31} & 0.6 \\ \hline
\multirow{12}{*}{\textbf{Spectrogram}} & \multicolumn{1}{c|}{\textbf{spectro}} & \textbf{VGG} & \multicolumn{1}{c|}{0.68} & \multicolumn{1}{c|}{0.72} & \multicolumn{1}{c|}{0.5} & \multicolumn{1}{c|}{0.75} & \multicolumn{1}{c|}{0.73} & 0.69 \\
 & \multicolumn{1}{c|}{\textbf{Log-Mel}} & \textbf{VGG} & \multicolumn{1}{c|}{0.74} & \multicolumn{1}{c|}{0.54} & \multicolumn{1}{c|}{0.79} & \multicolumn{1}{c|}{0.77} & \multicolumn{1}{c|}{0.73} & 0.71 \\
 & \multicolumn{1}{c|}{\textbf{MFCC}} & \textbf{VGG} & \multicolumn{1}{c|}{0.76} & \multicolumn{1}{c|}{0.58} & \multicolumn{1}{c|}{0.79} & \multicolumn{1}{c|}{0.83} & \multicolumn{1}{c|}{0.66} & 0.72 \\ \cline{2-9} 
 & \multicolumn{1}{c|}{\textbf{spectrogram}} & \textbf{ResNet} & \multicolumn{1}{c|}{0.67} & \multicolumn{1}{c|}{0.48} & \multicolumn{1}{c|}{0.74} & \multicolumn{1}{c|}{0.66} & \multicolumn{1}{c|}{0.78} & 0.67 \\
 & \multicolumn{1}{c|}{\textbf{Log-Mel}} & \textbf{ResNet} & \multicolumn{1}{c|}{0.6} & \multicolumn{1}{c|}{0.48} & \multicolumn{1}{c|}{0.64} & \multicolumn{1}{c|}{0.57} & \multicolumn{1}{c|}{0.76} & 0.61 \\
 & \multicolumn{1}{c|}{\textbf{MFCC}} & \textbf{ResNet} & \multicolumn{1}{c|}{0.76} & \multicolumn{1}{c|}{0.58} & \multicolumn{1}{c|}{0.77} & \multicolumn{1}{c|}{0.85} & \multicolumn{1}{c|}{0.73} & 0.74 \\ \cline{2-9} 
 & \multicolumn{1}{c|}{\textbf{spectrogram}} & \textbf{AlexNet} & \multicolumn{1}{c|}{0.7} & \multicolumn{1}{c|}{0.48} & \multicolumn{1}{c|}{0.75} & \multicolumn{1}{c|}{0.72} & \multicolumn{1}{c|}{0.56} & 0.66 \\
 & \multicolumn{1}{c|}{\textbf{Mel}} & \textbf{AlexNet} & \multicolumn{1}{c|}{0.74} & \multicolumn{1}{c|}{0.51} & \multicolumn{1}{c|}{0.79} & \multicolumn{1}{c|}{0.77} & \multicolumn{1}{c|}{0.71} & 0.7 \\
 & \multicolumn{1}{c|}{\textbf{MFCC}} & \textbf{AlexNet} & \multicolumn{1}{c|}{0.76} & \multicolumn{1}{c|}{0.56} & \multicolumn{1}{c|}{0.76} & \multicolumn{1}{c|}{0.78} & \multicolumn{1}{c|}{0.45} & 0.66 \\ \cline{2-9} 
 & \multicolumn{1}{c|}{\textbf{spectrogram}} & \textbf{LSTM} & \multicolumn{1}{c|}{0.73} & \multicolumn{1}{c|}{0.66} & \multicolumn{1}{c|}{0.75} & \multicolumn{1}{c|}{0.65} & \multicolumn{1}{c|}{0.58} & 0.67 \\
 & \multicolumn{1}{c|}{\textbf{Log-Mel}} & \textbf{LSTM} & \multicolumn{1}{c|}{0.75} & \multicolumn{1}{c|}{0.69} & \multicolumn{1}{c|}{0.77} & \multicolumn{1}{c|}{0.69} & \multicolumn{1}{c|}{0.65} & 0.71 \\
 & \multicolumn{1}{c|}{\textbf{MFCC}} & \textbf{LSTM} & \multicolumn{1}{c|}{0.77} & \multicolumn{1}{c|}{0.71} & \multicolumn{1}{c|}{0.79} & \multicolumn{1}{c|}{0.74} & \multicolumn{1}{c|}{0.74} & 0.75 \\ \hline
\multirow{3}{*}{\textbf{Transfer learning}} & \multicolumn{2}{c|}{\textbf{Wav2Vec}} & \multicolumn{1}{c|}{\textbf{0.90}} & \multicolumn{1}{c|}{\textbf{0.88}} & \multicolumn{1}{c|}{\textbf{0.88}} & \multicolumn{1}{c|}{0.88} & \multicolumn{1}{c|}{\textbf{0.89}} & \textbf{0.89} \\
 & \multicolumn{2}{c|}{\textbf{HuBERT}} & \multicolumn{1}{c|}{0.89} & \multicolumn{1}{c|}{0.82} & \multicolumn{1}{c|}{0.85} & \multicolumn{1}{c|}{\textbf{0.9}} & \multicolumn{1}{c|}{0.82} & 0.86 \\
 & \multicolumn{2}{c|}{\textbf{VGGish}} & \multicolumn{1}{c|}{0.83} & \multicolumn{1}{c|}{0.75} & \multicolumn{1}{c|}{0.8} & \multicolumn{1}{c|}{0.8} & \multicolumn{1}{c|}{0.77} & 0.79 \\ \hline

\end{tabular}
\end{table*}

\heading{Tabular features} Among the classifiers operating on tabular features, CatBoost showed the best performance with both feature sets (GeMAPS and ComParE) with overall AUC (of 0.72 and 0.6) respectively.

\heading{Spectrograms} Turning to spectrogram-based classifiers and comparing the three spectrograms (standard spectrogram, log-Mel, and MFCC), MFCC appears to capture more useful and discriminative information compared to other spectrogram types, leading to the model achieving superior performance.
The most powerful model is CNN-LSTM followed by ResNet50 VGG19 and Alexnet with AUC using MFCC as an input (0.75, 0.74, 0.72, 0.7) respectively, which means for classifying between the different BS patterns, it is more useful to involve temporal dynamics or sequential dependencies in addition to spatial patterns.
Moreover, CNN-LSTM showed more reliability in classifying the classes with the smallest sample sizes such as SB (AUC-0.71) using MFCC as an input, in comparison to VGG19, ResNet50, and AlexNet with AUC (0.58, 0.58, 0.56) using the same input. 

\heading{Transfer learning} Similar to the binary classification, using pre-trained models demonstrates the best performance across all the other models with the highest AUC achieved by Wav2Vec2.0 followed by HuBERT and VGGish (0.89, 0.86, 0.79) respectively.
It is noteworthy that all the pre-trained models showed reliable performance with all the classes regardless of the class sample size, which is highly effective in addressing challenges associated with small datasets.
This observation is clearly illustrated in Figure~\ref{fig6}, which presents the AUC values for each class achieved by the best models within each group. Even for the best-performing models, a significant drop in performance can be observed in tabular feature-based and spectrogram-based models for classes with smaller sample sizes (e.g., SB and HS) compared to those with larger sample sizes (e.g., None, MB, CRS). However, this limitation is effectively resolved by using pre-trained models.

\section{Conclusion}
Detecting bowel sound activity and patterns has been a significant challenge over the last century. The lack of high-quality data further complicates this task for traditional machine learning models. This work demonstrated the feasibility of differentiating different kinds of bowel sound patterns and provided a comparative assessment of different machine-learning approaches, ranging from decision trees on tabular expert features, over CNNs on spectrums to leveraging models pre-trained on large audio datasets. The best-performing models originated from the latter category, emphasizing the promising role of pre-trained models to overcome the challenges associated with small datasets. These advances pave the way for a better understanding of the value of bowel sound monitoring in gastrointestinal examinations, in particular for further studies to explore how these patterns correlate with gastrointestinal tract diseases.

While the raw dataset underlying this study cannot be shared due to consent restrictions, the full implementation of our approach, along with training scripts, is available in a corresponding code repository \url{https://github.com/AI4HealthUOL/bowel-sound-classification}.

\section*{Acknowledgments}
The authors acknowledge intramural funding from Faculty VI, Carl von Ossiezky Universität Oldenburg (Forschungspool, Potentialbereich mHealth) and from MWK Niedersachsen (via Fraunhofer IDMT, institute part HSA, project Connected Health).

\section*{References}
\bibliographystyle{IEEEtran}
\bibliography{bibfile}

\end{document}